\documentclass[12pt]{article}
\pdfoutput=1
\usepackage{epsfig}
\usepackage{amsmath}
\usepackage{color}
\usepackage{booktabs}
\usepackage{amssymb}
\usepackage{subfigure}

\usepackage{setspace}
\usepackage{array}
\usepackage{times}
\usepackage{bm}
\usepackage{graphicx}
\usepackage{latexsym}

\newtheorem{theorem}{Theorem}

\newtheorem{lemma}{Lemma}

\newtheorem{remark}{Remark}
\newtheorem{definition}{Definition}

\usepackage[dvips, letterpaper, nohead, top=1.3in, bottom=1.3in, left=1.3in, right=1.3in]{geometry}

\begin{document}

\small\normalsize

\title
{Adaptive Controls of FWER and FDR Under Block Dependence}

\author{
Wenge Guo\\
Department of Mathematical Sciences\\
New Jersey Institute of Technology\\
Newark, NJ 07102, U.S.A. \\
Email: wenge.guo@njit.edu
\and
Sanat Sarkar\\
Department of Statistics, Temple University\\
Philadelphia, PA 19122, U.S.A.
}

\maketitle

\begin{abstract}
Often in  multiple testing, the hypotheses appear in non-overlapping
blocks with the associated $p$-values exhibiting dependence within
but not between blocks. We consider adapting the Benjamini-Hochberg
method for controlling the false discovery rate (FDR) and the
Bonferroni method for controlling the familywise error rate (FWER)
to such dependence structure without losing their ultimate controls
over the FDR and FWER, respectively, in a non-asymptotic setting. We
present variants of conventional adaptive Benjamini-Hochberg and
Bonferroni  methods with proofs of their respective controls over
the FDR and FWER. Numerical evidence is presented to show that these
new adaptive methods can capture the present dependence structure
more effectively than the corresponding conventional adaptive
methods. This paper offers a solution to the open problem of
constructing adaptive FDR and FWER controlling methods under
dependence in a non-asymptotic setting and providing real
improvements over the corresponding non-adaptive ones.
\end{abstract}


\section{Introduction}

In many multiple hypothesis testing problems arising in modern
scientific investigations, the hypotheses appear in non-overlapping
blocks. Such block formation is often a natural phenomenon due to
the underlying experimental process or can be created based on other
considerations. For instance, the hypotheses corresponding to (i)
the different time-points in a microarray time-course experiment
(Guo, Sarkar and Peddada, 2010; Sun and Wei, 2011) for each gene; or
(ii) the phenotypes (or the genetic models) with (or using) which
each marker is tested in a genome-wide association study (Lei et
al., 2006); or (iii) the conditions (or subjects) considered for
each voxel in brain imaging (Heller et al. 2007), naturally form a
block. While applying multiple testing in astronomical transient
source detection from nightly telescopic image consisting of large
number of pixels (each corresponding to a hypotheses), Clements,
Sarkar and Guo (2012) considered grouping the pixels into blocks of
equal size based on telescope `point spread function.'

A special type of dependence, which we call block dependence, is the
relevant dependence structure that one should take into account
while constructing multiple testing procedures in presence of such
blocks. This dependence can be simply described by saying that the
hypotheses or the corresponding $p$-values are mostly dependent
within but not between blocks. Also known as the clumpy dependence
(Storey, 2003), this has been considered mainly in simulation
studies to investigate how multiple testing procedures proposed
under independence continue to perform under it (Benjamini, Krieger
and Yekutieli, 2006; Finner, Dickhaus, and Roters, 2007; Sarkar, Guo
and Finner, 2012, and Storey, Taylor and Siegmund, 2004), not in
offering FDR or FWER controlling procedures precisely utilizing it.
In this article, we focus on constructing  procedures controlling
the FDR and the FWER that incorporate the block dependence in a
non-asymptotic setting in an attempt to improve the corresponding
procedures that ignore this structure. More specifically, we
consider the Benjamini-Hochberg (BH, 1995) method for the FDR
control and the Bonferroni method for the FWER control and adapt
them to the data in two ways - incorporating the block dependence
and estimating the number of true null hypotheses capturing such
dependence.

Adapting to unknown number of true nulls has been a popular way to
improve the FDR and FWER controls of the BH and Bonferroni methods,
respectively. However, construction of such adaptive methods with
proven control of the ultimate FDR or FWER in a non-asymptotic
setting and providing real improvements under dependence is an open
problem (Benjamini, Krieger and Yekutieli, 2006; Blanchard and
Roquaine, 2009). We offer some solutions to this open problem in
this paper under a commonly encountered type of dependence, the
block dependence.

\section{Preliminaries}

Suppose that $H_{ij}$, $i=1, \ldots, b$; $j=1, \ldots, s_i$, are the
$n=\sum_{i=1}^b s_i$ null hypotheses appearing in $b$ blocks of size
$s_i$ for the $i$th block that are to be simultaneously tested based
on their respective $p$-values $P_{ij}$, $i=1, \ldots, b$; $j=1,
\ldots, s_i$. Let $n_0$ of these null hypotheses be true, which for
notational convenience will often be identified by $\hat P_{ij}$'s.
We assume that $\hat{P}_{ij} \sim U(0,1)$ and make the following
assumption regarding dependence of $P_{ij}$'s:

\vskip 10pt {\sc Assumption 1.} (Block Dependence) {\it The rows of
$p$-values $(P_{i1}, \ldots, P_{is_i})$, $i = 1, \ldots, b$, forming
the $b$ blocks are independent of each other.}

\vskip 10pt \noindent Under this assumption, the null $p$-values are
independent between but not within blocks. Regarding dependence
within blocks, our assumption will depend on whether we want to
control the FDR or FWER. More specifically, we develop methods
adapting to this block dependence structure and controlling the FDR
under positive dependence of the $p$-values within each block or the
FWER under arbitrary dependence of the $p$-values within each block.
The positive dependence condition, when assumed for each $i$, will
be of the type characterized by the following: \begin {eqnarray} E
\left \{ \phi_i(P_{i1}, \ldots, P_{is_i}) ~|~\hat P_{ij} \le u
\right \} \uparrow u \in (0,1), \end {eqnarray} for each $\hat
P_{ij}$ and any (coordinatewise) non-decreasing function $\phi_i$.
This type of positive dependence is commonly encountered and used in
multiple testing; see, for instance, Sarkar (2008) for references.
We will sometimes refer to block dependence more specifically as
{\it positive block dependence} in case when this dependence defined
by (1) in each block or as {\it arbitrary block dependence} in case
of any dependence within each block, to avoid any apparent double
meaning.

\vskip 10pt We will be using two types of multiple testing procedure
in this paper - stepup and single-step. Let $(P_i, H_i)$, $i=1,
\ldots, n$, be the pairs of $p$-value and the corresponding null
hypothesis, and $P_{(1)} \le \cdots \le P_{(n)}$ be the ordered
$p$-values. Given a set of critical constants $0 \le \alpha_1 \le
\cdots \le \alpha_n \le 1$, a stepup test rejects $H_i$ for all $i$
such that $P_i \le P_{(R)}$, where $R = \max \{1 \le i \le n:
P_{(i)} \le \alpha_i \}$, provided this maximum exists, otherwise,
it accepts all the null hypotheses. A single-step test rejects $H_i$
if $P_i \le c$ for some constant $c \in (0,1)$.

Let $V$ be the number of falsely rejected among all the $R$ rejected
null hypotheses in a multiple testing procedure. Then, the FDR or
FWER of this procedure, defined respectively by FDR = $E(V /\max
\{R,1 \})$ or FWER = ${\rm pr} (V \ge 1)$, is said to be controlled
at level $\alpha$, strongly unless stated otherwise, if it is
bounded above by $\alpha$. That is, for for any configuration of
true and false null hypotheses, the FDR or FWER of this procedure is
less than or equal to $\alpha$.

The BH method controlling the FDR at level $\alpha$ is a stepup test
with the critical constants $\alpha_i= i \alpha/n$; whereas, the
Bonferroni method controlling the FWER at level $\alpha$ is a
single-step test with the critical constant $\alpha/n$.

\section{Adaptive FDR control under block dependence}

The method we propose in this section is based on the idea of adapting
the BH method to the block dependence structure without losing the
ultimate control over the FDR in a non-asymptotic setting. Our adaptation
is done in two steps. First, we adjust it to
the block dependence structure and then develop its oracle version
given the number of true nulls.  Second, we consider the
data-adaptive version of this oracle method by estimating $n_0$
using an estimate that also captures the block dependence.

Towards adjusting the BH method to the block structure, we note that
it is natural to first identify blocks that are significant by
applying the BH method to simultaneously test the intersection null
hypotheses $\tilde H_i = \bigcap_{j=1}^{s_i} H_{ij}$, $i=1, \ldots,
b$, based on some block specific $p$-values, and then go back to
each significant block to see which hypotheses in that block are
significant. Let $\tilde P_i$, $i=1, \ldots, b$, be the block
$p$-values obtained by combining the $p$-values in each block
through a combination function. Regarding the choice of this
combination function, we note that the combination test for $\tilde
H_i$ based on $\tilde P_i$ must allow simultaneous testing of the
individual hypotheses $H_{ij}$, $j=1, \ldots, s_i$, with a strong
control of the FWER. This limits our choice to the Bonferroni
adjusted minimum $p$-value; see also Guo, Sarkar and Peddada (2010).
With these in mind, we consider adjusting the BH method as follows:

\begin {definition}
\emph{(Two-stage BH under block dependence)}
\begin {itemize}
\item []{\it 1. Choose $\tilde P_i = {\bar s}
\min_{1 \le j \le s_i} P_{ij}$ as the $i$th block $p$-value, for
$i=1, \ldots, b$, with ${\bar s} = \frac{1}{b} \sum_{i=1}^b s_i =
n/b$ being the average block size.} \

\item []{\it 2. Order the block $p$-values as $\tilde P_{(1)}
\le \cdots \le \tilde P_{(b)}$, and find $B = \max \{1 \le i \le b:
\tilde P_{(i)} \le i \alpha/b \}$.} \

\item []{\it 3. Reject $H_{ij}$ for all $(i,j)$ such that
$\tilde P_{i} \le \tilde P_{(B)}$ and $ P_{ij} \le B \alpha/n$,
provided the above maximum exists, otherwise, accept all the null
hypotheses.}
\end {itemize}
\end {definition}

The number of false rejections in this two-stage BH method is given by
\begin {eqnarray}
V = \sum_{i=1}^b \sum_{j=1}^{s_i} I(H_{ij}=0, P_{ij} \le B
\alpha/n), \nonumber
\end {eqnarray} where $H_{ij}= 0$ or $1$ according to whether it is true or false.
So, with $R$ as the total number of rejections, the FDR of this
method under block dependence is
\begin {eqnarray} {\rm FDR} &
= & \sum_{i=1}^b \sum_{j=1}^{s_i} I(H_{ij}=0)E \left ( \frac {I(
P_{ij} \le B \alpha/n)}{\max \{R, 1\}} \right ) \nonumber \\ & \le &
\sum_{i=1}^b \sum_{j=1}^{s_i} I(H_{ij}=0)E \left ( \frac {I(P_{ij}
\le B \alpha/n)}{\max \{B, 1\}} \right ), \end {eqnarray} since $R
\ge B$. For each $(i,j)$,
\begin {eqnarray}
\frac {I(P_{ij} \le B \alpha/n)}{\max \{B, 1\}} = \sum_{k=1}^b
\frac{I(P_{ij} \le k \alpha/n, B^{(-i)} = k-1)}{k},
\end {eqnarray}
where $B^{(-i)}$ is the number of significant blocks detected by the
adjusted BH method based on $\{\tilde P_1, \ldots, \tilde P_{b} \}
\setminus \{\tilde P_{i} \}$, the $b-1$ block $p$-values other than
the $\tilde P_i$, and the critical values $ i \alpha/b$, $i=2,
\ldots, b$. Taking expectation in (3) under the block dependence and
applying it to (2), we see that \begin {eqnarray} {\rm FDR} & \le &
\frac{\alpha}{n} \sum_{i=1}^b \sum_{j=1}^{s_i} I(H_{ij}=0)
\sum_{k=1}^b {\rm pr} (B^{(-i)} = k-1) \nonumber \\ & = &
\frac{\alpha}{n} \sum_{i=1}^b \sum_{j=1}^{s_i} I(H_{ij}=0) = \pi_0
\alpha, \end {eqnarray} where $\pi_0 = n_0/n$. Thus, we have the
following result holds:

\vskip 5pt

{\sc Result 1.} {\it The above defined two-stage BH method strongly
controls the FDR at $\alpha$ under Assumption 1 of arbitrary block
dependence.}

\vskip 5pt

If $n_0$, and hence $\pi_0$, were known, the FDR control of this
two-stage BH method could be made tighter, from $\pi_0\alpha$ to
$\alpha$, by shrinking each $p$-value from $P_{ij}$ to $\pi_0
P_{ij}$. This would be the oracle form of the adjusted BH method.
Since $\pi_0$ is unknown, one would consider using $\widehat \pi_0$
to estimate $\pi_0$ from the available $p$-values and then use the
estimate $\widehat \pi_0$ to define the so-called shrunken or
adaptive $p$-values $Q_{ij}=\widehat \pi_0 P_{ij}$ to be used in
place of the original $p$-values in the adjusted BH method. This
will be our proposed adaptive BH method.

For estimating $n_0$ capturing the block dependence structure before
defining the adaptive $p$-values, we consider using an estimate of
the form $\widehat n_0(\mathbf{P})$ that satisfies the following
property. In this property, ${\mathbf{P}}= ((P_{ij}))$ denotes the
set of $p$-values and ${\mathbf{H}} = ((H_{ij}))$.

\vskip 10pt  {\sc Property 1.} {\it  Let $\widehat n_0(\mathbf{P})$
be a non-decreasing function of each $P_{ij}$ such that \begin
{eqnarray} \sum_{i=1}^{b}\sum_{j=1}^{s_i}I(H_{ij} = 0) E_{DU} \left
\{ \frac{1}{\widehat n_0(\mathbf{P}^{(-i)}, \mathbf{0})} \right \}
\le 1, \end {eqnarray} where $\mathbf{P}^{(-i)}$ is the subset of
$p$-values obtained by deleting the $i$th row, $\hat
n_0({\mathbf{P}}^{(-i)}, {\mathbf{0}})$ is obtained from $\hat n_0
({\mathbf{P}})$ by replacing the entries in the $i$th row of
$\mathbf{P}$ by zeros, and $E_{DU}$ is the expectation under the
Dirac-uniform configuration of $\mathbf{P}^{(-i)}$, that is, when
the $p$-values in $\mathbf{P}^{(-i)}$ that correspond to the false
null hypotheses are set to $0$ and each of the remaining $p$-values
are considered to be uniformly distributed on [0,1]. }

We are now ready to define our proposed adaptive BH method in the following:

\begin {definition}
\emph{(Adaptive BH under block dependence)}
\begin {itemize}
\item []{\it 1. Consider an estimate $\widehat n_{0}(\mathbf{P})$ satisfying
Property 1 and define the adaptive $p$-values $Q_{ij} = \widehat \pi_0 P_{ij}$ using $\widehat \pi_0 = \widehat n_0/n$.} \

\item []{\it 2. Find $B^{*} = \max \{1 \le i \le b: \tilde Q_{(i)} \le i \alpha/ b \}$, where $\tilde Q_{(i)} = \widehat \pi_0 \tilde P_{(i)}$.} \

\item []{\it 3. Reject $H_{ij}$ for all $(i,j)$ such that $\tilde Q_{i} \le
\tilde Q_{(B^{*})}$ and $ Q_{ij} \le B^{*} \alpha/n$, provided the
maximum in Step 2 exists, otherwise, accept all the null
hypotheses.}
\end {itemize}
\end {definition}

\begin {theorem} Consider the block dependence structure in which the $p$-values are
positively dependent as in (1) within each block. The FDR of the
above adaptive BH method is controlled at $\alpha$ under such
positive block dependence.
\end {theorem}

A proof of this theorem will be given in Appendix.

What is exactly an estimate satisfying Property 1 that one can use
in this adaptive BH? The following result, which is again going to
be proved in Appendix, provides an answer to this question.

\vskip 5pt

{\sc Result 2.} {\it Consider the estimate  \begin {eqnarray}
\widehat n_0^{(1)} = \frac {n - R(\lambda) + s_{\max}}{1-\lambda},
\end {eqnarray} for any $ \left (2b+3 \right )^{-\frac{2}{b+2}} \le
\lambda <1$, where $s_{\max} = \max_{1 \le i \le b} s_i$ and
$R(\lambda) = \sum_{i=1}^b \sum_{j=1}^{s_i}I(P_{ij} \le \lambda)$ is
the number of $p$-values in $\mathbf{P}$ not exceeding $\lambda$. It
satisfies Property 1 under Assumption 1.}

\vskip 5pt

Based on Theorem 1 and Result 2, we have the following result.

\vskip 5pt

{\sc Result 3.} {\it The adaptive BH method of the above type based
on the estimates $\widehat n_0^{(1)}$ with $ \left (2b+3 \right
)^{-\frac{2}{b+2}} \le \lambda <1$ strongly controls the FDR at
$\alpha$ under the positive block dependence considered in Theorem
1.}

\vskip 5pt

\begin {remark} \rm When $s_{\max}=1$, the estimate $\widehat n_0^{(1)}$
reduces to $$\widehat n_0^{(0)} = \frac {n - R(\lambda) +
1}{1-\lambda},$$ the Storey at al.'s (2004) estimate, considered in
the context of adaptive FDR control (by Benjamini, Krieger and
Yekutieli, 2006; Blanchard and Roquain, 2009; Sarkar, 2008 and
Storey, Taylor and Siegmund, 2004) without any block structure. Of
course, Result 1 holds for any $\lambda \in (0,1)$ when
$s_{\max}=1$. Also, in this case, we are basically assuming that the
$p$-values are independent. Thus, as a special case, Result 3
provides the following known result available in the aforementioned
papers:

\vskip 5pt

{\sc Note 1.} {\it The adaptive BH method in Definition 2 based on
the estimate $\widehat n_0^{(0)}$ controls the FDR at $\alpha$ under
independence of the $p$-values. } \vskip 5pt
\end {remark}

\begin {remark} \rm Blanchard and Roquain (2009) presented an
adaptive BH method that continues to control the FDR under the same
dependence assumption of the $p$-values as made for the original BH
method. Their idea is to estimate $n_0$ independently through an
FWER controlling method before incorporating that into the original
BH method. While this adaptive BH method would be applicable to our
present context, it does not capture the group structure of the
data. Moreover, their simulation studies only show an improvement of
their adaptive BH method over the original BH method in very limited
situations. Hu, Zhao and Zhou (2010) considered adjusting the BH
method in presence of group structure by weighting the $p$-values
according to the relative importance of each group before proposing
its adaptive version by estimating these weights. However, this
version of the adaptive BH method is known to control the FDR only
in an asymptotic setting and under weak dependence.  \end {remark}

\section{Adaptive FWER control under block dependence}

Our proposed method here is based on the idea of adapting the
Bonferroni method to the block dependence  structure with ultimate
control of the FWER in a non-asymptotic setting. Given an estimate
$\hat n_0$ of $n_0$ obtained from the available $p$-values, the
Bonferroni method can be adapted to the data through $\hat n_0$ by
rejecting $H_{ij}$ if $P_{ij}\le \alpha/\hat n_0$; see, for
instance, Finner and Gontscharuk (2009) and Guo (2009). Our method
is such an adaptive version of the Bonferroni method, but based on
an estimate of $n_0$ satisfying Property 1 that captures the block
dependence.

\begin{definition}
\emph{(Adaptive Bonferroni under block dependence)}

{\it 1. Define an estimate $\widehat n_{0}(\mathbf{P})$ satisfying
Property 1.

2. Reject $H_{ij}$ if $P_{ij} \le \alpha/\widehat
n_{0}(\mathbf{P})$.}

\end{definition}

\begin {theorem} Consider the block dependence structure in which the
$p$-values within each block are  arbitrarily dependent. The FWER of
the above adaptive Bonferroni method is controlled at
$\alpha$ under such arbitrary block dependence.
\end {theorem}

This will be proved in Appendix. Based on Theorem 2 and Result 2, we
have the following result.

\vskip 5pt

{\sc Result 4.} {\it The adaptive Bonferroni method of the above
type based on the estimates $\widehat n_0^{(1)}$ with $ \left (2b+3
\right )^{-\frac{2}{b+2}} \le \lambda <1$ strongly controls the FWER
at $\alpha$ under the arbitrary block dependence considered in
Theorem 2.}

\vskip 5pt

\begin {remark} \rm Similar to what we have said in Remark 1 on
adaptive FDR control, the Storey at al.'s (2004) estimate $\widehat
n_0^{(0)}$ corresponding to the case $s_{\max}=1$ was also
considered in the context of adaptive FWER control (by Finner and
Gontscharuk, 2009, Guo, 2009, and Sarkar, Guo and Finner, 2012), of
course without any block structure. Thus, as a special case of
Result 4, we get the following result derived in these papers:

\vskip 5pt

{\sc Note 2.} {\it The adaptive Bonferroni method in Definition 3
based on the estimate $\widehat n_0^{(0)}$ controls the FWER at
$\alpha$ under independence of the $p$-values.} \vskip 5pt

\end {remark}

\section {Simulation studies}

We performed simulation studies to investigate the following
questions:

\begin {itemize}
\item[Q1.] How does the newly suggested adaptive BH method
based on the estimate $\widehat n_0^{(1)}$ perform in terms of the
FDR control and power with respect to the block size $s$, the
parameter $\lambda$, and the strength of dependence among the
$p$-values compared to the original BH method and the two existing
adaptive BH methods in Storey et al. (2004) and Benjamini et al.
(2006)?
\item[Q2.] How does the newly suggested adaptive Bonferroni
method based on the estimate $\widehat n_0^{(1)}$ perform in terms
of the FWER control and power with respect to the block size $s$,
the parameter $\lambda$, and the strength of dependence among the
$p$-values compared to the original Bonferroni method and the
existing adaptive Bonferroni method based on the estimate $\widehat
n_0^{(0)}$?
\end {itemize}

To simulate the values of FDR (or FWER) and average power, the
expected proportion of false nulls that are rejected, for each of
the methods referred to in Q1 and Q2, we first generated $n$ block
dependent normal random variables $N(\mu_i, 1), i = 1, \ldots, n$,
with $n_0$ of the $\mu_i$'s being equal to $0$ and the rest being
equal to $d =\sqrt{10}$, and a correlation matrix $\Gamma =
I_{\frac{n}{s}} \otimes \left [ (1-\rho)I_s+  \rho1_s 1_s^{\prime}
\right ]$ with the block size $s$ and non-negative correlation
coefficient $\rho$ within each block. We then applied each method to
the generated data to test $H_i: \mu_i = 0$ against $K_i: \mu_i \neq
0$ simultaneously for $i =1, \ldots, n$, at level $\alpha = 0.05$.
We repeated the above two steps $2,000$ times.

In the simulations on adaptive BH methods, we set $n = 240, n_0 =
120, s = 2, 3, 4,$ or $6$ and $\lambda = 0.2, 0.5,$ or $0.8$. Thus,
$n=240$ block dependent normal random variables $N(\mu_i, 1), i = 1,
\ldots, n$, are generated and grouped into $b=120, 80, 60,$ or $40$
blocks. When $s=2, 4,$ or $6$, half of the $\mu_i$'s in each block
are $0$ while the rest are $d =\sqrt{10}$. When $s=3$, one $\mu_i$
is $0$ while the rest are $d=\sqrt{10}$ in each of the first $40$
blocks, and two $\mu_i$'s are $0$ while the rest are $d=\sqrt{10}$
in each of the remaining $40$ blocks. Similarly, in the simulations
on adaptive Bonferroni methods, we set $n = 100, n_0 = 50$, $s = 2,
4, 10,$ or $20,$ and $\lambda = 0.2, 0.5,$ or $0.8$. Thus, $n=100$
block dependent normal random variables $N(\mu_i, 1), i = 1, \ldots,
n$, are generated and grouped into $b=50, 25, 10,$ or $5$ blocks,
with half of the $\mu_i$'s in each block being $0$ while the rest $d
=\sqrt{10}$.

The following are the observations from the above simulations:

{\it From Figure 1 and 2}: The simulated FDRs and average powers for
the three adaptive BH methods remain unchanged with increasing
$\rho$ for different values of $s$ and $\lambda$. For small $s$ and
different $\lambda$, all these three adaptive BH methods seem to be
more powerful than the conventional BH method. However, when $s$ is
large, the new adaptive method seems to lose its edge over the
conventional BH method.

\begin{figure}
\begin{center}
\includegraphics[scale = 0.6]{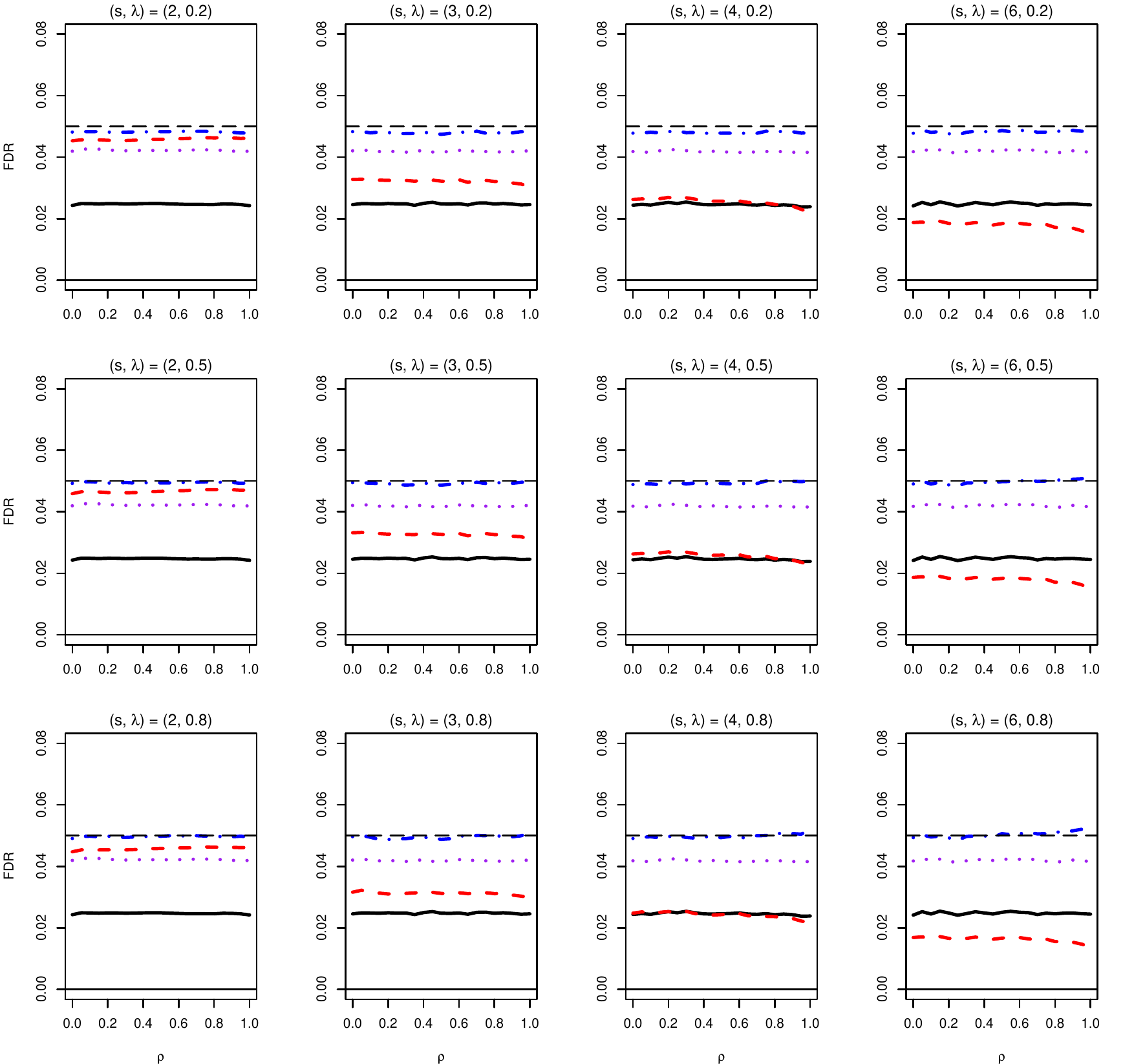}
\end{center}
\caption{
Simulated FDRs of the four multiple testing methods -- the original
BH and the three adaptive BH methods (adBH1, based on $\widehat
n_0^{(0)}$; adBH2, based on $\widehat n_0^{(1)}$; adBH3, the
adaptive BH method introduced in Benjamini et al, 2006) with $n =
240, n_0 = 120$, $s = 2, 3, 4$ or $6$, and $\lambda=0.2, 0.5$ or
$0.8$ at level $\alpha = 0.05$. [BH -- solid; adBH1 -- dot-dashes;
adBH2 -- long dashes; adBH3 -- dotted.]}
\end{figure}

\begin{figure}
\begin{center}
\includegraphics[scale = 0.6]{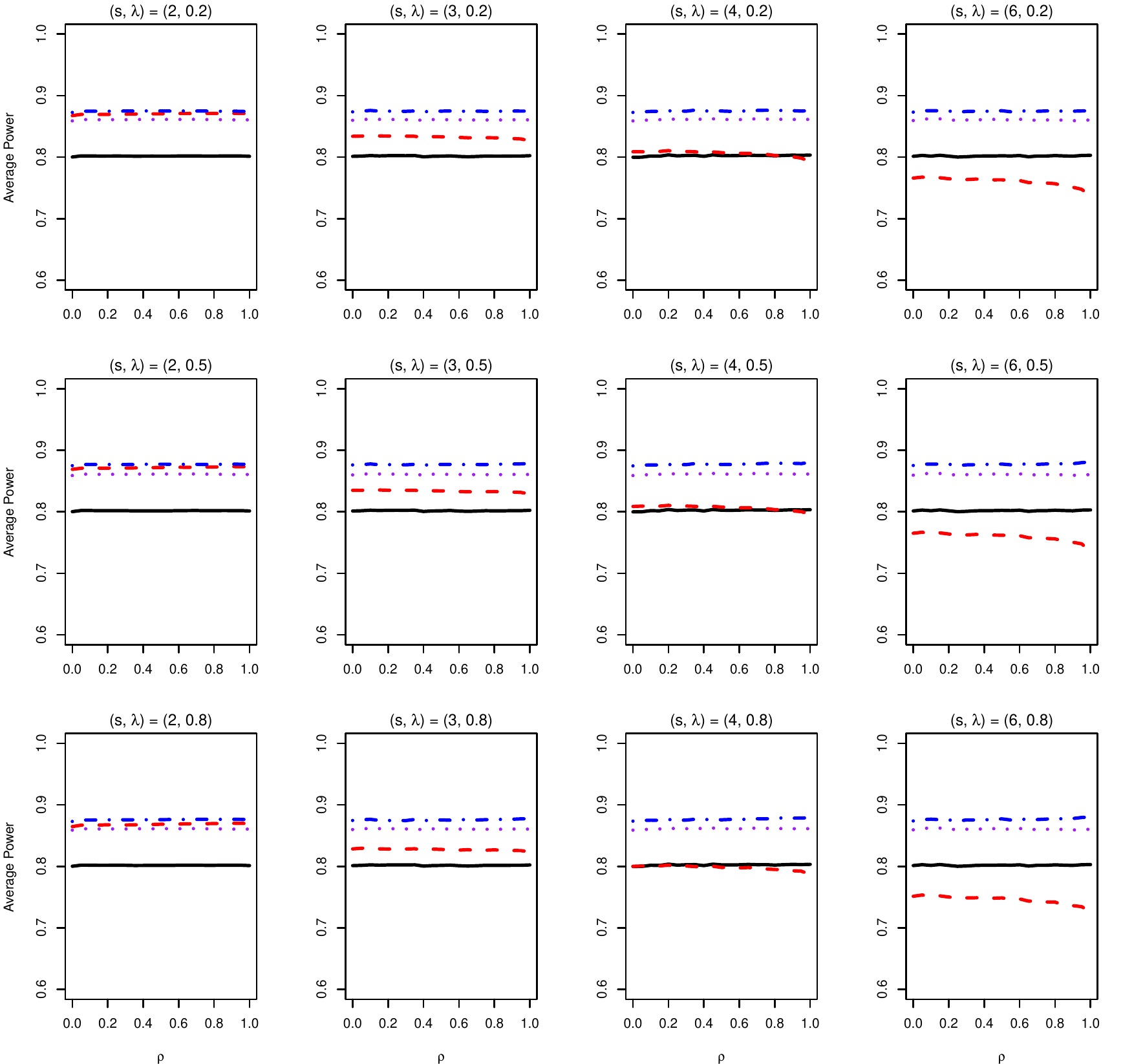}
\end{center}
\caption{
Average powers of the four multiple testing methods -- the original
BH and the three adaptive BH methods (adBH1, based on $\widehat
n_0^{(0)}$; adBH2, based on $\widehat n_0^{(1)}$; adBH3, the
adaptive BH method introduced in Benjamini et al, 2006) with $n =
240, n_0 = 120$, $s = 2, 3, 4$ or $6$, and $\lambda=0.2, 0.5$ or
$0.8$ at level $\alpha = 0.05$. [BH -- solid; adBH1 -- dot-dashes;
adBH2 -- long dashes; adBH3 -- dotted.]}
\end{figure}

\begin{figure}\label{fig1}
\begin{center}
\includegraphics[scale = 0.6]{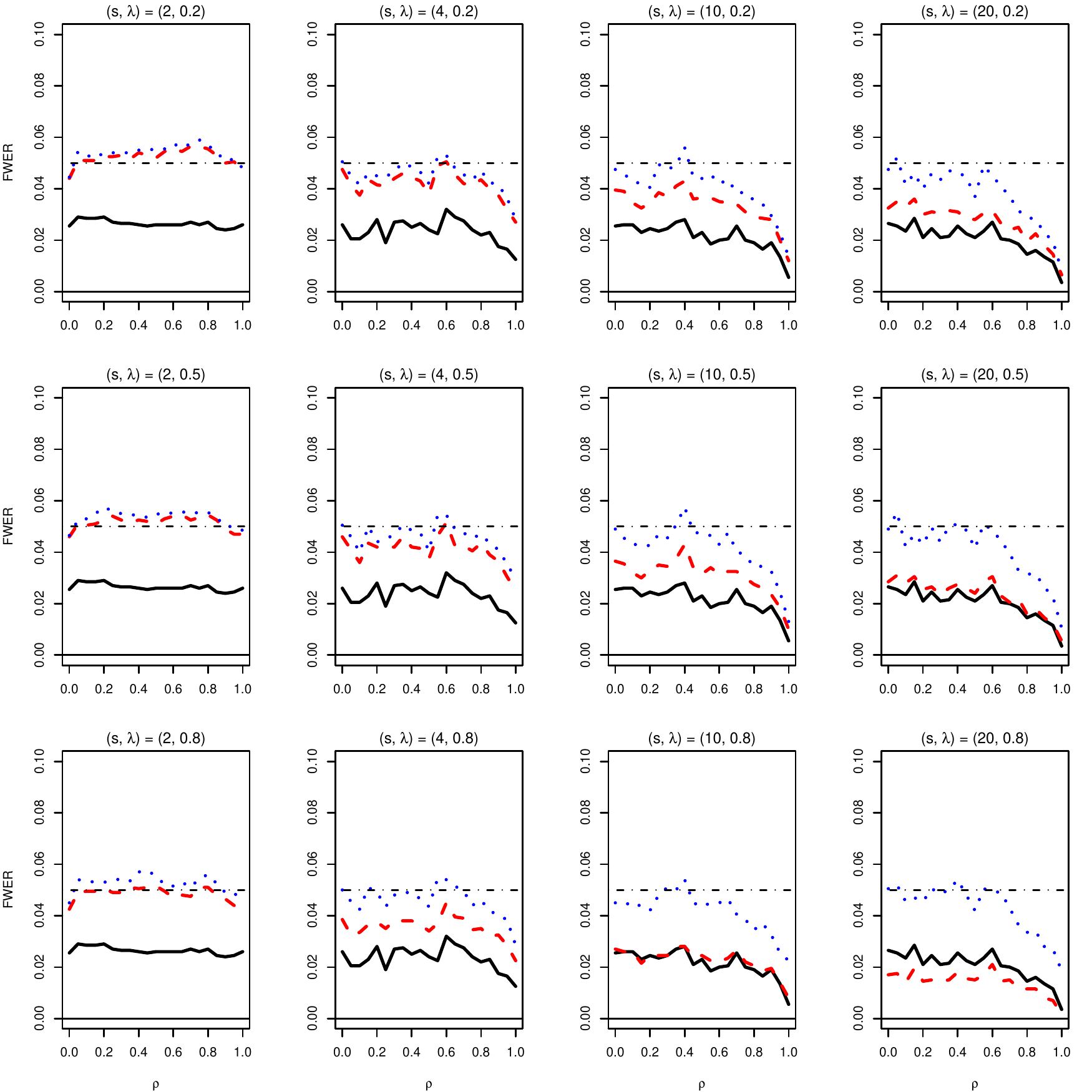}
\end{center}
\caption{Simulated FWERs of the three multiple testing methods --
the original Bonferroni method (Bonf.) and the two adaptive
Bonferroni methods (adBon1, based on $\widehat n_0^{(0)}$; adBon2,
based on $\widehat n_0^{(1)}$) with $n = 100, n_0 = 50$, $s = 2, 4,
10$ or $20$, and $\lambda=0.2, 0.5$ or $0.8$ at level $\alpha =
0.05$. [Bonf -- solid; adBon1 -- dotted; adBon2 -- long dashes]}
\end{figure}

\begin{figure}\label{fig2}
\begin{center}
\includegraphics[scale = 0.6]{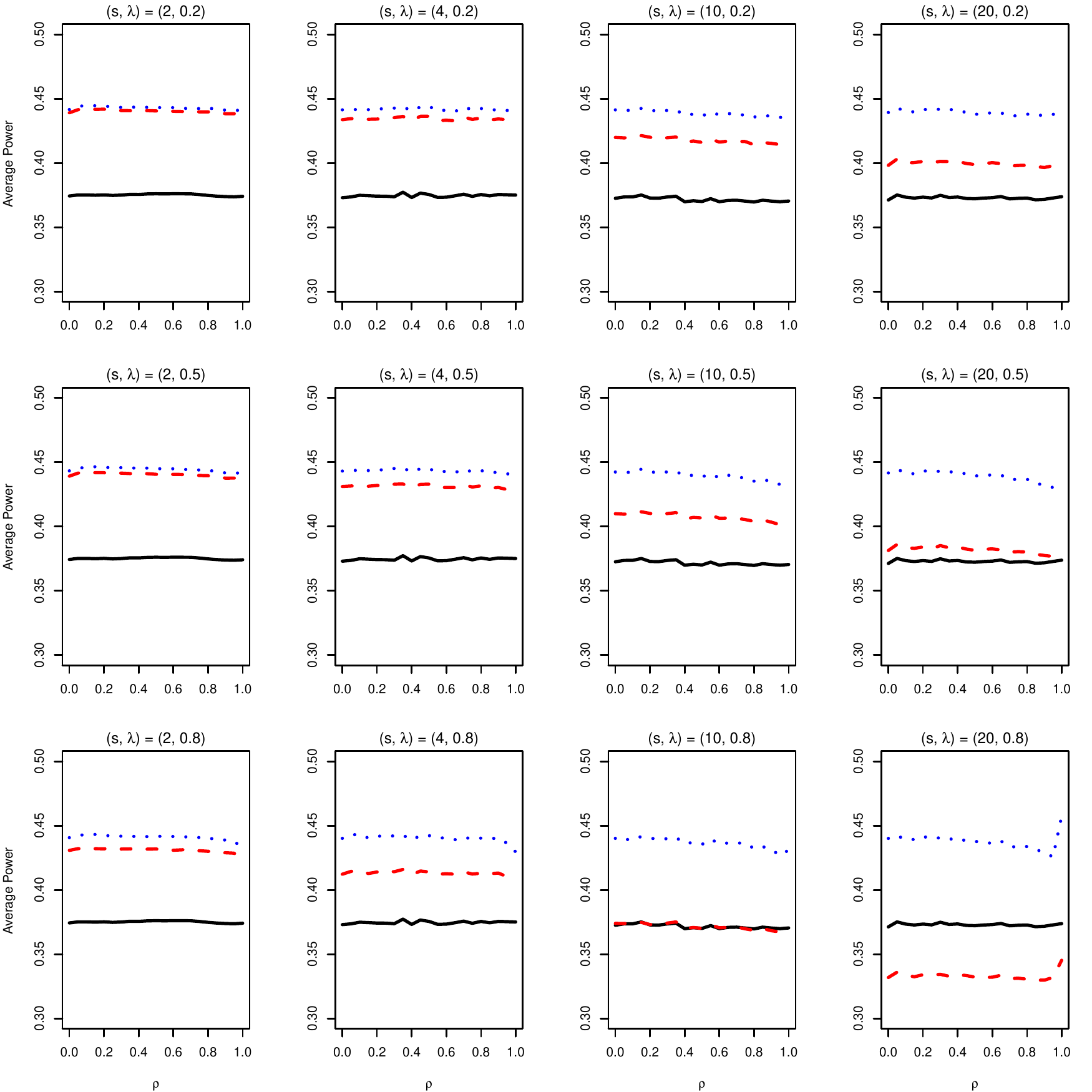}
\end{center}
\caption{Average
powers of the three multiple testing methods -- the original
Bonferroni method (Bonf.) and the two adaptive Bonferroni methods
(adBon1, based on $\widehat n_0^{(0)}$; adBon2, based on $\widehat
n_0^{(1)}$) with $n = 100, n_0 = 50$, $s = 2, 4, 10$ or $20$, and
$\lambda=0.2, 0.5$ or $0.8$ at level $\alpha = 0.05$. [Bonf --
solid; adBon1 -- dotted; adBon2 -- long dashes]}
\end{figure}

{\it From Figure 3 and 4}: When $s$ and $\lambda$ are both small,
both adaptive Bonferroni methods slightly lose the control over the
FWER for most values of $\rho$; however, when $\lambda$ is chosen to
be large, the FWER of the new adaptive method is controlled at
$\alpha$ with increasing $\rho$, whereas the existing adaptive
method still loses  control of the FWER. When $s$ is moderate or
large, the new adaptive method maintains a control over the FWER
whatever be the $\rho$, whereas the existing adaptive method can
lose control over the FWER for some values of $\rho$. In addition,
comparing the power performances of the two adaptive methods along
with their FWER control, it is clear that the new method is a better
choice as an adaptive version of the Bonferroni method under block
dependence than the existing one when $s$ is not very large.
However, when $s$ is very large, the new method loses its edge over
the existing one.

\section{Concluding remarks}

Construction of adaptive multiple testing methods with proven
control of the ultimate FDR or FWER under dependence in
non-asymptotic setting is an open problem. In this paper, we have
offered a solution to this open problem under a commonly assumed
form of dependence, the block dependence. We have developed new
adaptive BH method with proven FDR control under positive block
dependence and new adaptive Bonferroni method with proven FWER
control under arbitrary block dependence. They often provide real
improvements over the corresponding conventional BH and Bonferroni
methods.

The type of block dependence structure we consider here is often seen in real applications.
It perfectly fits in genetic research where the locations are independent
on different chromosomes but dependent inside the same chromosome. It also arises in the
context of simultaneous testing of multiple families of hypotheses, which is
often considered in large scale data analysis in modern scientific investigations, such as
DNA microarray and fMRI studies. Each family of null hypotheses here can be
interpreted as a block.

Benjamini and Bogomolov (2014) recently discussed a related problem of testing
multiple families of hypotheses and developed a related procedure:
Use the BH procedure across families, and then use the Bonferroni
procedure within the selected families, with the $B/b$ adjustment,
where $B$ is the number of the selected families and $b$ is the
number of the tested families. However, in the aforementioned paper,
the objective is to control a general average error rate over the
selected families including average FDR and FWER instead of the
overall FDR and FWER, which is different from ours. Also, there is
no explicit discussions of adaptive procedures in that paper as in
the methods suggested in this paper. It would be interesting to
investigate the connection between the theory and methods developed
in this paper and those in aforementioned paper.

\section*{Acknowledgements}
The research of the first author is supported by the NSF Grants
DMS-1006021, DMS-1309162 and the research of the second author is
supported by the NSF Grant DMS-1006344, DMS-1309273.

\section*{Appendix}

\noindent {\sc Proof of Theorem 1.} Proceeding as in showing [in (4)] that the FDR of the
adjusted BH method is bounded above by $\pi_0\alpha$, we first have
\begin {eqnarray} {\rm FDR} & \le
& \sum_{i=1}^b \sum_{j=1}^{s_i} I(H_{ij}=0) \sum_{k=1}^b \frac{{\rm
pr}(Q_{ij} \le k \alpha/n, {B^{*}}^{(-i)} =
k-1)}{k}, \end {eqnarray}  where ${B^{*}}^{(-i)}$ is the number of
significant blocks detected by the BH method based on the $b-1$
block specific adaptive $p$-values $\{\tilde Q_1, \ldots, \tilde Q_{b} \} \setminus
\{\tilde Q_{i} \}$ and the critical values $ i \alpha/b$,
$i=2, \ldots, b$. For each $(i,j)$,
\begin {eqnarray}
& & \frac {1}{k} I(H_{ij}=0 ) \sum_{k=1}^b {\rm pr}\left (Q_{ij} \le k \alpha/n,
{B^{*}}^{(-i)} = k-1 \right ) \nonumber \\ &
= & \frac {1}{k} I(H_{ij}=0 ) \sum_{k=1}^b {\rm pr}\left (P_{ij} \le
k \alpha/\widehat n_0(\mathbf{P}), {B^{*}}^{(-i)} = k-1
\right ) \nonumber \\
& \le & \frac {1}{k} I(H_{ij}=0 ) \sum_{k=1}^b {\rm pr}\left (
P_{ij} \le k \alpha/\widehat n_0(\mathbf{P}^{(-i)}, \mathbf{0}),
{B^{*}}^{(-i)} = k-1 \right ) \nonumber \\
& \le & \alpha E \left \{ \frac{I(H_{ij}=0 )} {\widehat n_0
(\mathbf{P}^{(-i)}, \mathbf{0} )} \sum_{k=1}^b {\rm pr} \left
({B^{*}}^{(-i)} = k-1~\big |~ P_{ij} \le k \alpha/\widehat
n_0(\mathbf{P}^{(-i)}, \mathbf{0}), \mathbf{P}^{(-i)} \right )
\right \}.  \nonumber \\
& &
\end {eqnarray}

Now,
\begin {eqnarray} & & \sum_{k=1}^b {\rm pr} \left ({B^{*}}^{(-i)}
= k-1~\big |~ P_{ij} \le k \alpha/\widehat n_0(\mathbf{P}^{(-i)},
\mathbf{0}), \mathbf{P}^{(-i)} \right ) \nonumber \\ & = &
\sum_{k=1}^b {\rm pr} \left ({B^{*}}^{(-i)} \ge k -1 ~\big |~ P_{ij}
\le k \alpha/\widehat n_0(\mathbf{P}^{(-i)}, \mathbf{0}),
\mathbf{P}^{(-i)} \right ) - \nonumber \\ & & \qquad
\sum_{k=1}^{b-1} {\rm pr} \left ({B^{*}}^{(-i)} \ge k ~\big |~
P_{ij} \le k \alpha/\widehat n_0(\mathbf{P}^{(-i)}, \mathbf{0}),
\mathbf{P}^{(-i)} \right ) \nonumber \\ & \le &  \sum_{k=1}^b {\rm
pr} \left ({B^{*}}^{(-i)} \ge k -1 ~\big |~ P_{ij} \le k
\alpha/\widehat n_0(\mathbf{P}^{(-i)}, \mathbf{0}),
\mathbf{P}^{(-i)} \right ) - \nonumber \\ & & \qquad
\sum_{k=1}^{b-1} {\rm pr} \left ({B^{*}}^{(-i)} \ge k ~\big |~
P_{ij} \le (k+1)\alpha/\widehat n_0(\mathbf{P}^{(-i)}, \mathbf{0}),
\mathbf{P}^{(-i)} \right ) \nonumber \\ & = & {\rm pr} \left
({B^{*}}^{(-i)} \ge 0 ~\big |~ P_{ij} \le \alpha/\widehat
n_0(\mathbf{P}^{(-i)}, \mathbf{0}), \mathbf{P}^{(-i)} \right ) = 1.
\end {eqnarray}
The validity of the inequality in (9) can be argued as follows:
Since $(P_{i1}, \ldots, P_{is_i})$ is independent of
$\mathbf{P}^{(-i)}$ and $I ({B^{*}}^{(-i)} \ge k )$ is decreasing in
$P_{ij}$'s, the conditional probability \[ {\rm pr} \left
({B^{*}}^{(-i)} \ge k ~\big |~ P_{ij} \le l \alpha /\widehat
n_0(\mathbf{P}^{(-i)}, \mathbf{0}), \mathbf{P}^{(-i)} \right ), \]
considered as a function of $l$, with $k$ and $\mathbf{P}^{(-i)}$
being fixed, is of the form
$$g(l) = E \left \{\phi \left (P_{i1},
\ldots, P_{is_i} \right )~|~P_{ij} \le l u \right \},$$ for a
decreasing function $\phi$ and a constant $u>0$. From the positive
dependence condition assumed in the theorem, we note that $g(l)$ is
decreasing in $l$, and hence $g(k+1) \le g(k)$.

From (7)-(9), we finally get \begin {eqnarray} {\rm FDR} \le
\alpha \sum_{i=1}^b \sum_{j=1}^{s_i} E \left \{ \frac{I(H_{ij}=0 )}
{\widehat n_0 (\mathbf{P}^{(-i)}, \mathbf{0} )} \right \} \le
\alpha, \end {eqnarray} which proves the desired result.  \hfill $\square$
\vskip 10pt

\noindent {\sc Proof of Result 2.} Before we proceed to prove this
result, we state two lemmas in the following that will facilitate
our proof. These lemmas will be proved later after we finish proving
the result. \vskip 10pt
\begin{lemma} Given a $p \times q$ matrix ${\mathbf{A}} =
((a_{ij}))$, where $a_{ij} = 0$ or $1$ and $\sum_{i=1}^p\sum_{j=1}^q
a_{ij} = m$, the entries of $\mathbf{A}$ can be always rearranged to
form a new $p \times q$ matrix ${\mathbf{B}} = ((b_{ij}))$ in such a
way that, for each $j=1, \ldots, q$, the entries in the $j$th column
of $\mathbf{B}$ are the entries of $\mathbf{A}$ in different rows,
$\sum_{i=1}^p b_{ij} = \lfloor \frac{m}{q} \rfloor$ or $\lfloor
\frac{m}{q} \rfloor +1$, and $\sum_{i=1}^p\sum_{j=1}^q b_{ij} = m$.
\end{lemma}

\begin {lemma} The function $f(x) = (2x+3)^{-\frac{2}{x+2}}$ is increasing
in $x \ge 1$ and $f(x) \le f(1)$ for all $0 \le x \le 1$. \end
{lemma}

We are now ready to prove the result. First, note that the result is
unaffected if we augment $\mathbf{P}$ to a complete $b \times
s_{\max}$ matrix by adding $s_{\max} - s_i$ more cells in the $i$th
row containing only $0$'s and assuming that the $H_{ij}$'s
corresponding to these additional zero $p$-values are all equal to
$1$, for each $i=1, \ldots, b$. In other words, we will assume
without any loss of generality, while proving this result, that
$\mathbf{P}$ is a $b \times s_{\max}$ matrix with $s_{\max}-s_i$
entries in the $i$th row being identically zero. Let $s_{\max} = s$
for notational convenience.

Consider the expectation
\begin {eqnarray} E_{DU} \left \{ \frac{1}{\widehat
n_0^{(1)}({\mathbf{P}}^{(-i)}, \mathbf{0})} \right \}, \nonumber
\end {eqnarray} in terms of $\mathbf{P}^{(-i)}$. Let
$\mathbf{H}^{(-i)}$ be the sub-matrix of $\mathbf{H}$ corresponding
to $\mathbf{P}^{(-i)}$. Since this expectation remains unchanged
under the type of rearrangements considered in Lemma 1 for
${\mathbf{H}}^{(-i)}$, we can assume without any loss of generality
that the number of true null $p$-values in the $j$th column of
${\mathbf{P}}^{(-i)}$ is $n_{0j}^{(-i)} = \lfloor \frac{n_0-m_i}{s}
\rfloor$ or $\lfloor \frac{n_0-m_i}{s} \rfloor +1$ for each $j=1,
\ldots, s$, where $m_i = \sum_{j=1}^s I(H_{ij} = 0)$.

Let $\widehat W_{j}^{(-i)}(\lambda) = \sum_{i^{'}(\neq i)=1}^b I
(H_{i^{'}j} = 0, P_{i^{\prime}j} > \lambda)$, for $j=1, \ldots, s$.
Under Assumption 1 and the Dirac-uniform configuration of
${\mathbf{P}}^{(-i)}$, $\widehat W_{j}^{(-i)}(\lambda) \sim {\rm
Bin} (n_{0j}^{(-i)}, 1 - \lambda)$. So, we have \begin {eqnarray} &
& E_{DU} \left \{ \frac{1}{\widehat n_0^{(1)}({\mathbf{P}}^{(-i)},
\mathbf{0})} \right \} =  E \left \{ \frac{1-\lambda}{\sum_{j=1}^s
\left [ \widehat W_{j}^{(-i)}(\lambda) + 1 \right ] } \right \}
\nonumber \\ & \le & \frac {1}{s^2} \sum_{j=1}^s E \left \{
\frac{1-\lambda}{\widehat W_{j}^{(-i)}(\lambda) + 1 } \right \} =
\frac {1}{s^2} \sum_{j=1}^s
\frac{1-\lambda^{n_{0j}^{(-i)}+1}}{n_{0j}^{(-i)} + 1}, \end
{eqnarray} with the first inequality following from the well-known
inequality between the arithmetic and harmonic means or using the
Jensen inequality and the second equality following from the result:
$E \left \{ (1+X)^{-1} \right \} = [1 -
(1-\theta)^{n+1}]/(n+1)\theta$, for $X \sim {\rm Bin} (n, \theta)$
(see, for instance, Liu and Sarkar, 2010).

Let $n_0 - m_i = (a_i + \beta_i)s$, for some non-negative integer
$a_i$ and $0 \le \beta_i < 1$. Note that
\begin {eqnarray}
a_is \le n_0 \le (a_i+\beta_i+1)s.
\end {eqnarray}
Also, $(1-\beta_i)$ proportion of the $s$ values $n_{0j}^{(-i)}$,
$j=1, \ldots, s$, are all equal to $a_i$ and the remaining $\beta_i$
proportion are all equal to $a_i+1$. So, the right-hand side of (11)
is equal to \begin {eqnarray} & & \frac {1}{s} \left [ \frac
{1-\beta_i}{a_i+1} \left ( 1 - \lambda^{a_i+1} \right ) + \frac
{\beta_i}{a_i+2} \left ( 1 - \lambda^{a_i+2} \right ) \right ] \le
\frac {1}{s} \left [ \frac {1-\beta_i}{a_i+1} + \frac
{\beta_i}{a_i+2} \right ] \left ( 1 - \lambda^{a_i+2} \right )
\nonumber \\ & = &  \frac {(a_i+2 -\beta_i)(1 -
\lambda^{a_i+2})}{s(a_i+1)(a_i+2)} \le \frac {(a_i+1+\beta_i) (a_i+2
-\beta_i)(1 - \lambda^{a_i+2})}{n_0(a_i+1)(a_i+2)} \nonumber \\ & =
& \frac{1}{n_0}\left [ 1 + \frac
{\beta_i(1-\beta_i)}{(a_i+1)(a_i+2)} \right ] \left (1 -
\lambda^{a_i+2} \right ) \le \frac{1}{n_0}\left [ 1 + \frac
{1}{4(a_i+1)(a_i+2)} \right ] \left (1 - \lambda^{a_i+2} \right ).
\nonumber \end {eqnarray} Here, the second inequality follows from
(12). The desired inequality (5) then holds for this estimate if
\begin {eqnarray} & & \left [ 1 + \frac {1}{4(a_i+1)(a_i+2)} \right
] \left (1 - \lambda^{a_i+2} \right ) \le 1, \nonumber \end
{eqnarray} which is true if and only if
\begin{eqnarray}
\lambda \ge \left [ 1+4(a_i+1)(a_i+2) \right ]^{-\frac{1}{a_i+2}} =
(2a_i+3)^{-\frac{2}{a_i+2}}.
\end {eqnarray}
Let $f(a_i) = (2a_i+3)^{-\frac{2}{a_i+2}}$. As seen from (12), $a_i
\le n_0/s \le b$,  thus, the inequality $f(b) \ge f(a_i)$ holds for
all $a_i \ge 0$, since $f(b) \ge f(a_i)$ if $a_i \ge 1$ and $f(b)
\ge f(1) \ge f(a_i)$ if $0 \le a_i \le 1$, due to Lemma 2. So, the
inequality (13) holds if $\lambda \ge \left (2b+3 \right
)^{-2/(b+2)}$. This completes our proof of Result 2.  \hfill $\square$

\vskip 10pt \noindent {\bf Proof of Lemma 1.} Let $s = (s_1, \ldots,
s_q)$ be the column sum vector of $\mathbf{A}$, that is, $s_j =
\sum_{i=1}^p a_{ij}, j = 1, \ldots, q$, and $\sum_{j=1}^q s_j = m$.
Without any loss of generality, we can assume that $s_1 \ge \ldots
\ge s_q$. Consider a given column sum vector $s^* = (s^*_1, \ldots,
s^*_q)$ satisfying $s^*_1 \ge \ldots \ge s^*_q$, where $s^*_j =
\lfloor \frac{m}{q} \rfloor$ or $\lfloor \frac{m}{q} \rfloor + 1$
for $j = 1, \ldots, q$, and $\sum_{j=1}^q s^*_j = m$.

We prove that $s^*$ is majorized by
$s$; that is, for each $k = 1, \ldots, q$,
\begin{equation}
\sum_{j=k}^q s^*_j \ge \sum_{j=k}^q s_j.
\end{equation}
Suppose the inequality (14) does not hold for some $k = 1, \dots,
q$. Let $k_1 = \max \{k: \sum_{j=k}^q s^*_j < \sum_{j=k}^q s_j \}$.
Since $s_{k_1} > s^*_{k_1}$, thus for each $j = 1, \ldots, k_1-1$,
$s_j \ge s_{k_1} \ge s^*_{k_1}+1 \ge \lfloor \frac{m}{q} \rfloor +1
\ge s_j^*$, implying that
\begin{eqnarray}
\sum_{j=1}^q s_j = \sum_{j=1}^{k_1 -1} s_j + \sum_{j=k_1}^q s_j >
\sum_{j=1}^{k_1 -1} s_j^* + \sum_{j=k_1}^q s^*_j = m, \nonumber \end {eqnarray}
which is a contradiction. So, $s^*$ is majorized by
$s$.

By Theorem 2.1 of Ryser (1957), one can rearrange the $1$'s in the
rows of $\mathbf{A}$ to construct a  new $p \times q$ matrix which
has the column sum vector $s^*$. Thus, the desired result follows.
 \hfill $\square$

\vskip 10pt \noindent {\bf Proof of Lemma 2.} Let $g(x) = \ln f(x) =
-\frac{2}{x+2} \ln (2x + 3)$ for $x \ge 0$ and $\varphi(u) = \ln u -
\frac{1}{u}-1$ for $u \ge 3$. Thus,
\begin{equation}
g'(x) = \frac{1}{(x+2)^2} \left [2\ln (2x + 3) -\frac{4x + 8}{2x+3}
\right] = \frac{2\varphi(2x+3)}{(x+2)^2}. \nonumber
\end{equation}
Note that $\varphi(u)$ is a strictly increasing continuous function
in $[3, \infty)$ with $\varphi(3) < 0$ and $\varphi(5) > 0$, thus
there exists a unique $u^* \in (3, 5)$ satisfying $\varphi(u^*)=0$.
Let $x^* = \frac{u^*-3}{2}$, then $x^* \in (0, 1)$ and $g'(x^*) =
0$. Thus, $g'(x) < 0$ for $x \in [0, x^*)$ and $g'(x) > 0$ for $x
\in (x^*, \infty)$. Based on $x^* < 1$, we have that $g'(x) > 0$ for
$x \ge 1$ and $g(x) \le \max\{g(0), g(1)\} = \max\{-\ln3, -2\ln5/3\}
= g(1)$ for $0 \le x \le 1$. Thus, the desired result follows.
 \hfill $\square$

\vskip 10pt
\noindent{\sc Proof of Theorem 2.} The FWER of the method in this theorem is given by
\begin {eqnarray}
{\rm FWER} & = & {\rm pr} \left \{ \bigcup_{i=1}^b
\bigcup_{j=1}^{s_i} \left (P_{ij} \le \frac{\alpha
I(H_{ij}=0)}{\widehat n_0(\mathbf{P})} \right ) \right \} \le
\sum_{i=1}^b \sum_{j=1}^{s_i} {\rm pr} \left \{ P_{ij} \le
\frac{\alpha I(H_{ij}=0)}{\widehat
n_0(\mathbf{P})} \right \} \nonumber \\
& \le & \sum_{i=1}^b \sum_{j=1}^{s_i}  {\rm pr} \left \{ P_{ij} \le
\frac{\alpha I(H_{ij}=0)}{\widehat n_{0}(\mathbf{P}^{(-i)},
\mathbf{0})}\right \} \le \alpha \sum_{i=1}^b \sum_{j=1}^{s_i}
E_{DU} \left \{
\frac{I(H_{ij}=0)}{\widehat n_0(\mathbf{P}^{(-i)}, \mathbf{0})} \right \}  \nonumber \\
& \le & \alpha. \end {eqnarray} In (15), the first inequality
follows from the Bonferroni inequality, the second and third follow
from the non-decreasing property of $\widehat n_0$ and that $\hat
P_{ij} \sim U(0, 1)$ and the assumption of arbitrary block
dependence, and the fourth follows from the condition (5) satisfied
by $\widehat n_0$. Thus, the desired result is proved.
 \hfill $\square$


\begin{thebibliography}{99}

\small

{

\bibitem{r40}
{\sc Benjamini, Y. \&\ Bogomolov, M.} (2014). Selective inference on
multiple families of hypotheses. \textit{J. Roy. Statist.
Soc. Ser. B} {\bf 76}, 297--318.


\bibitem{r30}
{\sc Benjamini, Y. \&\ Hochberg, Y.} (1995). Controlling the false
discovery rate: A practical and powerful approach to multiple
testing. \textit{J. Roy. Statist. Soc. Ser. B~} {\bf 57}, 289-300.


\bibitem{r1}
{\sc Benjamini, Y, Krieger, K. \&\ Yekutieli, D.} (2006).
\newblock Adaptive linear step-up procedures that control the false
discovery rate.
\newblock {\em Biometrika} {\bf 93}, 491-507.

\bibitem{by:2001}
{\sc Benjamini, Y. \&\ Yekutieli, D.} (2001).
\newblock The control of the false discovery rate in multiple testing under
  dependency.
\newblock \textit{Ann. Statist.} \textbf{29}, 1165--1188.

\bibitem{r2}
{\sc Blanchard, G. \&\ Roquain, E.} (2009).
\newblock Adaptive FDR control under independence
and dependence.
\newblock {\em J. Mach. Learn.} {\bf 10}, 2837--2871.


\bibitem{r3}
{\sc Clements, N., Sarkar, S. \&\ Guo, W.} (2012).
\newblock Astronomical transient detection using grouped p-values
and controlling the false discovery rate.
\newblock In \emph{Statistical Challenges in Modern Astronomy}, edited
by Eric D. Feigelson and G. Joseph Babu, Lecture Notes in Statistics,
Vol. 209, Part 4, Springer-Verlag, 383-396.

\bibitem{r31}
{\sc Finner, H., Dickhaus, T. \&\ Roters, M. } (2007).
\newblock Dependency and false discovery rate: Asymptotics.
\newblock {\em Ann. Statist.} {\bf 35}, 1432-1455.

\bibitem{r4}
{\sc Finner, H. \&\ Gontscharuk, V.} (2009).
\newblock Controlling the familywise error rate with plug-in
estimator for the proportion of true null hypotheses.
\newblock {\em J. Roy. Statist. Soc., Ser. B} {\bf 71}, 1031--1048.

\bibitem{r6}
{\sc Gavrilov, Y., Benjamini, Y. \&\ Sarkar, S. K.} (2009).
\newblock An adaptive step-down procedure with proven FDR control.
\newblock {\em Ann. Statist.} {\bf 37}, 619--629.

\bibitem{r7}
{\sc Guo, W.} (2009). \newblock A note on adaptive Bonferroni and
Holm procedures under dependence. \newblock {\em Biometrika}, {\bf
96}, 1012-–1018.

\bibitem{r8}
{\sc Guo, W., Sarkar, S. \&\ Peddada, S.} (2010).
\newblock Controlling false discoveries in multidimensional directional
decisions, with applications to gene expression data on ordered
categories.
\newblock {\em Biometrics } {\bf 66}, 485-492.


\bibitem{r81}
{\sc Heller R., Golland Y., Malach R. \&\ Benjamini Y.} (2007).
\newblock Conjunction Group Analysis: An alternative to mixed/random effect
analysis.
\newblock {\em NeuroImage} {\bf 37}, 1178-1185.

\bibitem{r9}
{\sc Hochberg, Y. \&\ Benjamini, Y.} (1990).
\newblock More powerful procedures for multiple significance testing.
\newblock {\em Statist. Med.} {\bf 9}, 811-–818.

\bibitem{r10}
{\sc Hochberg, Y., \&\ Tamhane, A.~ C.} (1987).
\newblock {\em Multiple Comparison Procedures}.
\newblock Wiley: New York.

\bibitem{r91}
{\sc Hu, J., Zhao, H. \&\ Zhou, H.} (2010).
\newblock False Discovery Rate Control with Groups.
\newblock {\em J. American Statistical Association} {\bf 105},
1215-1227.

\bibitem{r11}
{\sc Lei S., Chen Y., Xiong D., Li L. \&\ Deng H.} (2006)
\newblock Ethnic difference in osteoporosis-related phenotypes and its
potential underlying genetic determination.
\newblock {\em Journal of Musculoskelet Neuronal Interact.} {\bf 6}, 36-46.

\bibitem{r131}
{\sc Liu, F. \&\ Sarkar, S. K.} (2010).
\newblock A note on estimating the false discovery rate under mixture model.
\newblock {\em Journal of Statistical Planning and Inference} {\bf 140}, 1601-1609.

\bibitem{r13}
{\sc Liu, F. \&\ Sarkar, S. K.} (2011).
\newblock A new adaptive method to control the false discovery rate.
\newblock {\em Series in Biostatistics} {\bf 4}, World Scientific,
3-26.

\bibitem{r14}
{\sc Romano, J. P., Shaikh, A. M. \&\ Wolf, M.} (2008).
\newblock Control of the false discovery rate under dependence using the bootstrap and subsampling.
\newblock {\em TEST} {\bf 17}, 417--442.

\bibitem{r141}
{\sc Ryser, H. J. } (1957).
\newblock Combinatorial properties of matrices of zeros and ones.
\newblock {\em Canad. J. Math.} {\bf 9}, 371-377.

\bibitem{r15}
{\sc Sarkar, S.~K.} (1998).
\newblock Some probability inequalities for ordered MTP$_2$ random variables:
a proof of the Simes conjecture.
\newblock {\em Ann. Statist.} {\bf 26}, 494--504.

\bibitem{r16}
{\sc Sarkar, S.~K.} (2002).
\newblock Some results on false discovery rate in stepwise multiple testing procedures.
\newblock {\em Ann. Statist.} {\bf 30}, 239--257.

\bibitem{r17}
{\sc Sarkar, S. K.} (2006).
\newblock False discovery and false nondiscovery rates in single-step multiple
testing procedures.
\newblock {\em Ann. Statist.} {\bf 34}, 394 --415.

\bibitem{r19}
{\sc Sarkar, S. K.} (2008).
\newblock On methods controlling the false discovery rate (with discussions).
\newblock {\em Sankhya} {\bf 70}, 135--168.

\bibitem{r20}
{\sc Sarkar, S.~K. \&\ Chang, C-K.} (1997).
\newblock The Simes method for multiple hypothesis testing with positively dependent test statistics.
\newblock {\em J. Amer. Statist. Assoc.} {\bf 92}, 1601-1608.


\bibitem{r21}
{\sc Sarkar, S., Guo, W. \&\ Finner, H.} (2012).
\newblock On adaptive procedures controlling the familywise error rate.
\newblock {\em Journal of Statistical Planning and Inference}, \textbf{142}, 65-78.


\bibitem{r22}
{\sc Schweder, T. \&\ Spj$\phi$tvoll, E.} (1982).
\newblock Plots of p-values to evaluate many tests simulataneously.
\newblock {\em Biometrika} {\bf 69}, 493-–502.

\bibitem{r23}
{\sc Storey J.} (2003).
\newblock Comment on `Resampling-based multiple testing for DNA microarray
data analysis' by Ge, Dudoit, and Speed.
\newblock {\em Test } {\bf 12}, 52-60.

\bibitem{r24}
{\sc Storey, J.~D., Taylor, J.~E. \&\ Siegmund, D.} (2004).
\newblock Strong control, conservative point estimation and
simultaneous conservative consistency of false discovery rates: a
unified approach. \newblock {\em J. Roy. Statist. Soc., Ser. B} {\bf
66}, 187-205.

\bibitem{r25}
{\sc Sun W \&\ Wei Z.} (2011).
\newblock Multiple testing for pattern identification, with applications to
microarray time course experiments. \newblock {\em Journal of the
American Statistical Association} {\bf 106}, 73-88.

}


\end{thebibliography}
\end{document}